\documentclass[aps,prd,amsmath,amssymb,reprint,superscriptaddress,nofootinbib]{revtex4-2} 

\usepackage{amsmath}  
\usepackage{amsfonts} 
\usepackage{graphicx} 
\usepackage{color} 
\usepackage{amssymb}
\usepackage{appendix}
\usepackage{enumerate}

\begin{document}

\title{Gravitational Aharonov-Bohm Effect}

\author{RY Chiao}
\email{raymond\_chiao@yahoo.com}
\affiliation{University of California, Merced, School of Natural Sciences, P.O. Box 2039,
Merced, CA 95344, USA}
\author{NA Inan}
\email{ninan@ucmerced.edu}
\affiliation{Clovis Community College, 10309 N. Willow, Fresno, CA 93730 USA}
\affiliation{University of California, Merced, School of Natural Sciences, P.O. Box 2039,
Merced, CA 95344, USA}
\affiliation{Department of Physics, California State University Fresno, Fresno, CA 93740-8031, USA}
\author{M Scheibner}	\email{mscheibner@ucmerced.edu}
\affiliation{University of California, Merced, School of Natural Sciences, P.O. Box 2039,
Merced, CA 95344, USA}
\author{J Sharping}
\email{jsharping@ucmerced.edu}
\affiliation{University of California, Merced, School of Natural Sciences, P.O. Box 2039,
Merced, CA 95344, USA}
\author{DA Singleton}
\email{dougs@mail.fresnostate.edu}
\affiliation{Department of Physics, California State University Fresno, Fresno, CA 93740-8031, USA}
\author{ME Tobar}
\email{michael.tobar@uwa.edu.au}
\affiliation{Quantum Technologies and Dark Matter Labs, Department of Physics, University of Western Australia, Crawley, WA 6009, Australia.}

\date{\today}

\begin{abstract}
We investigate the gravitational  Aharonov-Bohm effect, by placing a quantum system in free-fall around a gravitating body {\it e.g.}  a satellite orbiting the Earth. Since the system is in free-fall, by the equivalence principle, the quantum system is locally in flat, gravity free space-time - it is screened from the gravitational field. For a slightly elliptical orbit the gravitational potential will change with time. This leads to the energy levels of the quantum system developing side bands which is the signature for this version of the Aharonov-Bohm effect. This contrasts with the normal signature of the Aharonov-Bohm effect of shifting of interference fringes.      
\end{abstract}

\maketitle

\newpage

\section{Introduction}

The original proposal for the Aharonov-Bohm (AB) effect \cite{AB} focused on the scalar and vector potentials of the electromagnetic interaction. In particular the seminal paper of Aharonov and Bohm \cite{AB} focused mostly on the AB effect connected with the vector potential and magnetic field (vector-magnetic AB effect) rather than the scalar potential and electric field (scalar-electric AB effect). The original experimental set-up for the scalar potential-electric field AB effect involved switching the potential on and off as the electric charges entered and exited metal tubes. These tubes acted as Faraday shells to shield the charges from the electric field but not from the electric scalar potential. Since the experimental set-up for the vector-magnetic AB effect is much easier to realize, there are many experimental tests of the vector-magnetic AB effect, beginning with the first experiments by Chambers \cite{chambers}, to the definitive, loophole-free experiments in the mid-1980s \cite{tono}, and through to the present. In contrast, the best test of the scalar-electric AB effect \cite{electric-ab} is not as clean, since it measures both scalar-electric and vector-magnetic effects together, and the charges are not completely shielded from electric fields.   

In reference \cite{chiao-2023} an alternative probe of the scalar-electric AB was proposed. In the standard, scalar-electric set-up, charges are sent along different paths which have a potential difference between them (but with the charges at all times shielded from the electric fields). The observational signature of the scalar-electric AB effect is a shift in the quantum interference pattern of charges. In contrast the proposal of reference  \cite{chiao-2023}  placed a quantum system (Rubidium atoms) inside a Faraday cage with a time varying scalar potential, $\Phi _e (t)$. (In \cite{chiao-2023}, $V(t)$ was used for the scalar potential.) The observational signature highlighted in \cite{chiao-2023} is the development of energy side-bands in the spectrum of the quantum system. Therefore, in this alternative approach, one has a shifting of energy levels as compared to a shifting of interference fringes of the standard set-up.

We will now apply the analysis of \cite{chiao-2023} to the \textit{gravitational} AB effect. By gravitational AB effect we mean the scalar-gravitational AB effect which depends on the Newtonian scalar potential. In contrast the vector-gravitational AB effect depends on the gravitational vector potential (which leads to the Lense-Thirring field). This vector-gravitational AB effect was discussed in \cite{chiao-2014}. There has been a recent experimental verification of the scalar-gravitational AB effect \cite{overstreet} which follows the standard procedure: split a matter beam into two paths, with one path experiencing a different gravitational potential compared to the other, and then observe a shift in the interference pattern when the beams are recombined. Here we apply the set-up for the scalar electric AB effect given in \cite{chiao-2023}, to the gravitational AB effect giving a cleaner confirmation of this effect. It is cleaner in the sense that the quantum system is in free fall and thus screened from the gravitational forces via the equivalence principle.

For the scalar-electric AB effect, the phase picked up by an electric charge is 
\begin{equation}
    \label{AB-elec}
    \varphi _e (t) = \frac{e}{\hbar} \int _0 ^t \Phi _e (t') dt' ~,
\end{equation}
where $\Phi _e (t')$ is the time dependent electric scalar potential. The subscript $e$ stands for electromagnetic interaction.  The gravitational version of \eqref{AB-elec} is
\begin{equation}
    \label{AB-grav}
    \varphi _g (t) = \frac{m}{\hbar} \int _0 ^t \Phi _g (t') dt' ~.
\end{equation}
The electric charge, $e$, has been replaced by gravitational ``charge", $m$ (mass), and the electric potential has been replaced by the gravitational potential, $\Phi _g$.

For the scalar-electric case the quantum system was placed inside a Faraday cage with a sinusoidal varying potential, $\Phi _e (t) = V_0 \cos (\Omega t)$. This set-up will not work for the gravitational case, for several reasons as we will discuss in the next two paragraphs. For the gravitational AB effect we will instead place our quantum system in a satellite in an almost circular, low Earth orbit. 

\section{Gravitational AB phase shift} -- For the gravitational AB effect we will consider a gravitational potential of the form
\begin{equation}
\Phi _g (  t )     =-\frac{G M }{r (t)}
\label{Ab-grav-2}
\end{equation}
where $G$ is Newton's constant, $M$ is the mass of some large body ({\it e.g.} the Earth) about which our quantum system will orbit, and  $r (t)$ is the time-dependent distance between the satellite and one focus of the orbit. Using \eqref{Ab-grav-2} in \eqref{AB-grav} one finds that the gravitational phase becomes
\begin{equation}
    \label{AB-grav-a}
    \varphi _g (t) = - \frac{m}{\hbar} \int _0 ^t \frac{G M }{r (t')} dt' ~.
\end{equation}
while the radius as a function of angle formula for a closed orbit is simple and well known ({\it i.e.} $r(\theta ) = \frac{r_0}{1-\epsilon \cos \theta}$ with $0 \le \epsilon < 1$ being the eccentricity) the radius as a function of time ({\it i.e.} $r(t)$) is not as simple or well known. Because of this we will focus on almost circular orbits, where the gravitational potential can be approximated as a simple oscillatory term plus a constant.  

This set-up is different from the scalar electric case where $\Phi _e (t)$ was varied by oscillating charges onto and off of a Faraday shell, thus giving a time dependent electric potential. In principle, such a method would work for the gravitational case, since oscillating charges on to and off of the shell, would also mean that one would be oscillating ``gravitational" charge ({\it i.e.} mass) on to and off of the shell. However, the gravitational interaction is so weak, and the amount of mass moved onto and off of the shell is so small, that the effect would be much too small to observe. We can counteract the smallness of the gravitational interaction, and smallness of the masses oscillated onto the shell, by instead using an astronomical, fixed mass $M$, and changing the potential by varying the distance of the quantum system with respect to the mass $M$ {\it i.e.} letting the distance of the quantum system from the central mass be time dependent $r(t)$. 

As mentioned above we will consider almost circular, low Earth orbits so that $r(t)$ can be approximated as a simple oscillatory term plus a constant. One might think using a highly elliptical orbits is preferable, since the change in gravitational potential between apogee and perigee would be larger. However, such highly elliptical orbits ({\it e.g.} Molniya orbits \cite{molniya}) have significantly longer periods and lower frequency, negating the advantage gained by the larger change in the gravitational potential. 

We now lay out the details of determining $r(t)$ for almost circular, low Earth orbits. The relevant parameter of such orbits are: (i) Perigee and apogee radius from the center of the Earth are $r_p = 6.800 \times 10^6$ m and $r_a = 6.810 \times 10^6$ m, respectively, which corresponds to a perigee altitude of 400 km and apogee altitude of 410 km given that the Earth's radius is $r_E \approx 6400$ km. These radii correspond roughly to those of the International Space Station. (ii) The period of a satellite with this apogee/perigee is about $T \approx$ 90 minutes or 5400 seconds giving an angular frequency of $\Omega = \frac{2 \pi}{T}  = 1.0 \times 10^{-3} \rm{\frac{rad}{sec}}$ (or $f= \frac{1}{T} = 1.85 \times 10^{-4} ~ {\rm Hz}$). The radius of the orbit as a function of time can be approximated as \footnote{This treatment of nearly circular orbits is essentially that found in section 9.5 of reference \cite{kk}}
\begin{equation}
    \label{radius}
    r(t) = \frac{r_p+r_a}{2} + \frac{r_p-r_a}{2} \cos(\Omega t) \equiv A + B \cos (\Omega t)~.
\end{equation}
Using the $r_p$ and $r_a$ values above, we find the $A$ and $B$ parameters defined in \eqref{radius} become $A=6.805 \times 10^6$ m and $B = -5.000 \times 10^3$ m. Perigee occurs at $t=0$ and apogee at $t=\pi/\Omega$. For the chosen $r_a$ and $r_p$, $A \gg B$ so one can approximate $\frac{1}{r(t)} = \frac{1}{A+B\cos (\Omega t)} \approx \frac{1}{A} \left(1 - \frac{B}{A} \cos (\Omega t ) \right)$. With this the gravitational potential in \eqref{Ab-grav-2} becomes
\begin{equation}
    \label{Ab-grav-4}
    \Phi _g (t) \approx  - \frac{GM}{A} \left[ 1 - \frac{B}{A} \cos (\Omega t ) \right]  ~,
\end{equation}
Inserting this in \eqref{AB-grav} gives 
\begin{eqnarray}
    \label{Ab-grav-3a}
    \varphi _g (t) &=& -\frac{GmM}{\hbar A} \int _0 ^t  \left( 1 - \frac{B}{A}  \cos (\Omega t') \right) dt'  \nonumber \\
    &=& -\frac{GmM}{\hbar A} t + \frac{GmMB}{\hbar \Omega A^2} \sin (\Omega t)  \\
    &\equiv& -\frac{GmM}{\hbar A} t + \alpha \sin (\Omega t) = -\frac{GmM}{\hbar A} t + \varphi ' _g  (t) ~. \nonumber 
\end{eqnarray} 
In the last line of \eqref{Ab-grav-3a} we have defined the dimensionless frequency modulation depth of modulation parameter $\alpha \equiv \frac{GmMB}{\hbar \Omega A^2}$.
It is the second sinusoidal term in \eqref{Ab-grav-3a} ({\it i.e.} the $\alpha \sin (\Omega t)$ term) which leads to the AB phase, and we have therefore split the phase $\varphi _g (t)$ into a linear term in $t$ and a sinusoidal term. As we will see below the term linear in $t$ can be packaged with the energy to give an overall shift of the base energy of the atomic system. It is the sinusoidal term, $\varphi ' _g (t)$ which gives the gravitational AB phase. We show this by solving the Schr{\"o}dinger equation for a quantum system (either atomic or nuclear) placed in the gravitational potential $\Phi _g (t)$.  

We carry out an analysis of the quantum system in the presence of this time varying gravitational potential. Since the quantum system is in free fall, it is effectively screened from the gravitational field and forces, which is required for AB effect experiments. 

In the absence of the gravitational potential, $\Phi _g (t)$, we assume that the quantum system has a known solution to the time-independent Schr{\"o}dinger ({\it i.e.} $H_0 \Psi _i ({\bf x}) = E_i \Psi_i ({\bf x})$) where $H_0$, $\Psi_i$ and $E_i$ are the Hamiltonian, wavefunction and energy eigenvalues, respectively, of the quantum system. The coordinate ${\bf x}$ is the relative (internal) coordinate for the quantum system {\it i.e.} the location of the electrons for an atomic system or the location of nucleons for nuclear systems. Placing the quantum system in the potential, $\Phi _g (t)$, leads to the Hamiltonian $H = H_0 + m \Phi _g (t)$, with the new term being the time-dependent gravitational potential energy. 

The solution to the Schr{\"o}dinger equation for this time-dependent gravitational Hamiltonian is found as in the scalar-electric case \cite{chiao-2023}. The Schr{\"o}dinger equation for $H$ is 
\begin{equation}
\label{tdse}
i\hbar \frac{\partial \psi }{\partial t}=H\psi =\left( H_{0}+m \Phi_g (t) \right) \psi.
\end{equation}
Now since we are in the frame of reference fixed with the quantum system the kinetic energy part of $H_0$ in \eqref{tdse} will be modified due to transforming from an inertial frame (where $H_0 \Psi _i ({\bf x}) = E_i \Psi_i ({\bf x})$ holds) to a orbiting, non-inertial frame. The relationship between the velocities in the inertial and orbiting frames is ${\bf v}_{in} = {\bf v}_{orb} + \Omega \times {\bf r} (t)$ \cite{kk} with $\Omega \times {\bf r} (t)$ being the orbital velocity of the satellite, and ${\bf r} (t)$ being the time-dependent position of the satellite. Due to the time dependence of the coordinate ${\bf r} (t)$ this additional term, $\Omega \times {\bf r} (t)$, will also generate a phase factor through a shift in the kinetic energy term in $H_0$, that is in addition to the phase factor that comes from the added potential term, $m \Phi _g (t)$, in \eqref{tdse}. To estimate this additional phase shift due to the change from inertial to orbiting frame we use the Virial theorem which shows that the average of the kinetic term associated with the orbital velocity, $\Omega \times {\bf r} (t)$, will have the opposite sign and be half the magnitude of the potential term {\it i.e.} $\langle T_{\Omega} \rangle = - \frac{1}{2} \langle V_g \rangle$. Thus the phase factor coming from the change to an orbiting frame will partially cancel the phase factor coming from the time changing gravitational potential. However, the overall phase factor will still be of the same order of magnitude and for simplicity we will focus on the phase factor coming from only $m \Phi_g (t)$.

To solve \eqref{tdse} we apply a separation-of-variables ansatz of the form
$\psi (\mathbf{x},t)=X(\mathbf{x})T(t)$ and substitute this into \eqref{tdse} to give
\begin{eqnarray}
i\hbar \frac{\partial \psi }{\partial t} &=& i\hbar X\frac{dT}{dt} = \left(
H_{0}+m\Phi_g\right) XT \nonumber \\
&=& TH_{0}X+X\left( m \Phi_g\right) T.
\end{eqnarray}
Dividing by $XT$ and moving $m\Phi_g (t)$ to the left hand side gives
\begin{equation}
-m\Phi_g + i\hbar \frac{1}{T}\frac{dT}{dt}=\frac{1}{X}H_{0}X.
\end{equation}
This equation has the form $f(t)=g\left( \mathbf{x}\right)$
where $f(t)$ is {\it only} a function of $t$, and $g\left( \mathbf{x}%
\right) $ is $only$ a function of $\mathbf{x}$. The only way that this can be true is if each function is equal to a constant: $f(t)=g\left( \mathbf{x}\right) =E$. This gives the separated equations
\begin{eqnarray}
-m \Phi _g +i\hbar \frac{d\ln T}{dt} = E ~~~~
\text{and}~~~~
H_{0}X = EX 
\label{temporal equation}
\end{eqnarray}%
Setting $X=\Psi _{i }({\bf x})$ and $E=E_{i }$, gives the time-independent Schr{\"o}dinger equation
$H_{0}\Psi _{i }({\bf x})=E_{i }\Psi _{i }({\bf x})$,
where $\Psi _i$ and $E_i$ are the wavefunction and eigen-energy, respectively, for the known eigenvalue problem of the unperturbed Hamiltonian, $H_0$. Integrating \eqref{temporal equation} over $t$ gives
\begin{equation}
-m\int \Phi _g (t) dt+i\hbar \int \frac{d\ln T (t)}{dt}dt=\int E_{i }dt \label{int-1}
\end{equation}
Carrying out the integration in \eqref{int-1} and solving for $T(t)$, gives  
\begin{eqnarray}
\label{T(t) solution}
T(t) &=&\exp \left( -\frac{i}{\hbar }E_{i }t\right) \exp \left( -\frac{i}{
\hbar }m\int \Phi_g dt\right)  \nonumber \\
&=&\exp \left( -\frac{i}{\hbar }\left(E_{i } +\frac{GmM}{A}\right) t-i\alpha \sin \Omega t\right) \\
&=&\exp \left( -\frac{i}{\hbar }\left(E_{i } + \frac{GmM}{A}\right)t-i\varphi ' _g (t)\right) \nonumber ~,  
\end{eqnarray}
For an atomic system, $m \to m_e = 9.11 \times 10^{-31}$ kg which is the electron mass. Using $M=5.97 \times 10^{24}$ kg for  the mass of the Earth, and inserting the values for $B, A$ and $\Omega$ given around equation \eqref{radius}, we find $\alpha_{atomic} \approx -3.7 \times 10^{11}$. For a nuclear system,
one has $m \to m_N =1.67 \times 10^{-27}$ kg. Using the same values for $M, A, B$ and $\Omega$, we find $\alpha_{nuclear} \approx -6.8 \times 10^{14}$.

The result in \eqref{T(t) solution} is similar to the scalar-electric AB result from \cite{chiao-2023}, in terms of the AB phase, $\varphi _g$, and the parameter $\alpha$. However the term with $\frac{GmM}{A}$ gives a constant shift to the unperturbed energy $E_i$ which was not present in \citep{chiao-2023}. This term presents a constant shift in the energy, $E_i$, due to the time-independent part of the gravitational potential. In the scalar-electric AB case, we were able to set this constant part of the electric potential equal to zero -- something not possible in the gravitational case. For the atomic case with $m=9.11 \times 10^{-31}$ kg, this shift is $\frac{GmM}{A} \approx 5.3 \times 10^{-23}$ J $\approx 3.3 \times 10^{-4}$ eV; for the nuclear case with  $m=1.67 \times 10^{-27}$ kg, this shift is $\frac{GmM}{A} \approx 9.8 \times 10^{-20}$ J $\approx 0.6$ eV. These constant shifts are small compared to usual atomic and nuclear energies of the unperturbed system.   
    
We put the above results together to obtain the wave function for $H=H_0 + m \Phi_g (t)$.  Multiplying $X({\bf x}) = \Psi _i ({\bf x})$ and $T(t)$ from \eqref{T(t) solution} gives the wave function, $\psi _i ({\bf r}, t)$ as 
\begin{equation}
\psi _{i }(\mathbf{r},t)=\Psi _{i }(\mathbf{r})\exp \left( -\frac{
i\left(E_{i }+\frac{GmM}{A}\right)t}{\hbar }-i\varphi ' _g (t)\right) ~. 
\label{psi}
\end{equation}
This new wave function is the original wave function with an added AB phase factor $\exp \left( -i\varphi ' _g (t)\right)$. Using \eqref{T(t) solution} gives $\varphi_g ' (t)$ as 
\begin{equation}
\varphi ' _g (t)=\frac{m}{\hbar }\int \frac{GMB}{A^2} \cos (\Omega t) dt  = \alpha \sin \Omega t ~, 
\label{phase modulation}
\end{equation}
Exponentiating $\varphi '_g (t)$ from \eqref{phase modulation} and using the Jacobi-Anger expansion gives
\begin{equation}
\begin{aligned}
\exp \left( - i\varphi ' _g (t)\right) &=\exp \left( - i\alpha \sin \Omega t\right) \\
&=\sum_{n=-\infty }^{\infty } (-1)^n J_{n}(\alpha )\exp \left( in\Omega t\right)~.
\end{aligned}
\label{Jacobi-Anger}
\end{equation}
Inserting the result from \eqref{Jacobi-Anger} back into \eqref{psi}, the wave function reads
\begin{equation}
\begin{aligned}
&\psi _i(\mathbf{r},t) =  \\
&\Psi _i(\mathbf{r})\sum_{n=-\infty}^{\infty } (-1)^n J_{n}(\alpha )\exp \left( in\Omega t\right) \exp \left( -\frac{i\left(E_i + \frac{GmM}{A} \right) t}{\hbar }\right)  \\
&=\Psi _i(\mathbf{r})\sum_{n=-\infty }^{\infty } (-1)^n J_{n}(\alpha )\exp\left( -\frac{i\left( E_i+\frac{GmM}{A}-n\hbar \Omega \right) t}{\hbar }\right)
\end{aligned}
\label{psi-2}
\end{equation}
Thus each energy level $E_i$ will be split into a multiplet $E_i^{(n)}$ with
\begin{equation}
E_i^{(n)}=E_i+ \frac{GmM}{A} \pm n\hbar \Omega  \equiv {\tilde E}_i \pm n\hbar \Omega ~,
\label{energy-2}
\end{equation}
where $n$ is an integer, and $E_i^{(n)}$ are evenly spaced energy levels, with an energy step $\hbar \Omega$. In the second equality, we have absorbed the small, constant gravitational shift $\frac{GmM}{A}$ into $E_i$ by defining ${\tilde E}_i$. This new energy spectrum, $E_i ^{(n)}$, is of the form of the quasi-energies discussed in \cite{zeldovich}. If one takes the results of equations \eqref{psi-2} and \eqref{energy-2} at face value, this would seem to imply a new spectrum with an infinite number of new states labeled by the side band index $n$. However, from \eqref{psi-2} one finds that the different contributions are weighed by the Bessel functions $J_n (\alpha)$. In Fig. 1, we plot $J_n (\alpha)$ for a wide range of $\alpha$'s as a function of $n$. For all values of $\alpha$, one finds rapid oscillation for $n < \alpha$. At $n \equiv n_{max} \approx |\alpha|$ there is a sharp up shoot, and for $n> n_{max} \approx \alpha$ $J_n (\alpha)$ exponentially decays to zero, so that states beyond $n_{max}$ do not contribute.
\begin{figure}[htb!]
\label{fig1}
    \centering
   \includegraphics[width=0.5\textwidth]{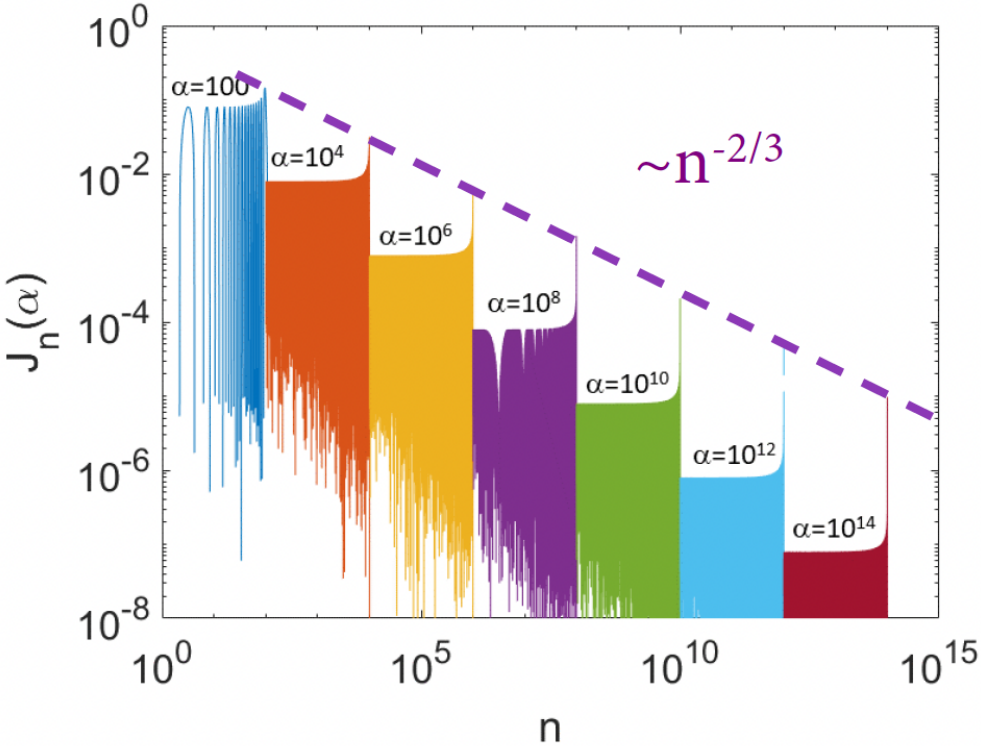}
    \caption{Weighting factor, $J_n (\alpha)$ versus $n$ from \eqref{psi-2} for various $\alpha$'s. Maximum $J_n (\alpha)$ occurs for $n=n_{max} \approx \alpha$. As $\alpha$ increases the maximum weighting factor, $J_\alpha (\alpha)$, decreases, but is sharper and relatively larger compared to the lower side bands. A power law fit of the maximum side band goes as $n^{-2/3}$.}
\end{figure}
Using the numerical values for $\alpha_{atomic}$ and $\alpha_{nuclear}$ given below \eqref{T(t) solution} gives $n_{max} \approx |\alpha _{atomic}| \approx 3.7 \times 10^{11}$, for atomic systems, and  $n_{max} \approx |\alpha _{nuclear}| \approx 6.8 \times 10^{14}$ for nuclear systems. Note that as $\alpha$ increases, the values of the maximum weighting, $J_\alpha (\alpha)$, decrease, but the up shoot at the maximum is sharper and larger, relative to weighting, $J_n (\alpha)$ for $n < n_{max} \approx \alpha$. 

The energy associated with the frequency, $\Omega$, of the nearly circular, low Earth orbits is $\hbar \Omega = 1.1 \times 10^{-37}$ J = $6.6 \times 10 ^{-19}$ eV. The split energy levels from \eqref{energy-2} for atomic systems is
\begin{equation}
E_i^{(n_{max})} = {\tilde E}_i \pm n_{max} \hbar \Omega \approx {\tilde E} _i \pm  2.4 \times 10^{-7} {\text eV},~~
\label{energy-3}
\end{equation}
while for nuclear systems \eqref{energy-2} gives
\begin{equation}
E_i^{(n_{max})} = {\tilde E}_i \pm n_{max} \hbar \Omega \approx {\tilde E} _i \pm  4.5 \times 10^{-4} {\text eV}.~~
\label{energy-4}
\end{equation}
Therefore the experimental signature of the gravitational AB effect is the observation of these side bands ($\pm 2.4 \times 10^{-7}$ eV for atomic systems, or $\pm 4.5 \times 10^{-4}$ eV for nuclear systems).

In \eqref{energy-3} and \eqref{energy-4} we have only included the maximum energy side bands. This is based on Fig. 1 which shows that the maximum side band, with $n=n_{max} \approx \alpha$, has a higher weighting, $J_\alpha (\alpha)$, relative to the weighting of the side bands, $J_n (\alpha )$, with $n < n_{max} \approx \alpha$. 

\section{Experimental set-ups for the gravitational AB effect} -- In order to observe the small shift in the energy side bands in \eqref{energy-3} and \eqref{energy-4}, we need to ensure that the widths of the transitions between the base energy level, ${\tilde E} _i$, are smaller than these values. Our atomic or nuclear systems need to have very narrow spectral lines - less than $\sim 10^{-7}$ eV  and $\sim 10^{-4}$ eV for atomic and nuclear systems, respectively. Also for both of the possible set-ups discussed below -- either atomic clocks  or M{\"o}ssbauer effect -- one needs to wait for some period of time (for example one orbital period) to allow the system to come to its steady state. The above theoretical analysis has been done under the assumption that the quantum system used to probe the gravitational AB effect has very long coherence times -- as is the case for the atomic clock system discussed next. In future work we will analyze the case when the quantum system have finite coherence times.

\subsection{Atomic system: Atomic clocks } -- A potential set-up to test for the  gravitationally induced side bands from \eqref{energy-3}  would use atomic clocks. We propose using the ACES (Atomic Clock Ensemble in Space) mission \cite{aces}, which will place optical/microwave frequency atomic clocks on the International Space Station (ISS). We need to determine if atomic clocks would be able to distinguish the side bands of order $10^{-7}$ eV as per \eqref{energy-3}. To this end we will use the two-sample variance (or Allan deviation), $\sigma _y ^2 (\tau)$, and the related spectral density, $S_y (f)$. These two quantities are described in detail in \cite{rutman}, but briefly for $S_y (f) = h_\alpha f^\alpha$, where $f$ is the frequency, and $h_\alpha$ is a measure of the noise level. For white noise, $\alpha =0$ so $S_y (f) = h_0$, and $\sigma _y ^2 (\tau) = \frac{h_0}{2 \tau}$, where $\tau$ is the time interval of the measurement. From \cite{laurent} the ACES cesium clocks have $\sigma _y = \frac{1.1 \times 10^{-13}}{\sqrt{\tau}}$. Using this and the expression  $\sigma _y ^2 (\tau) = \frac{h_0}{2 \tau}$, and\footnote{The averaging time, $\tau$, is taken as 1 second just as an illustration. A more realistic averaging time would be of the order of the orbital period of $\tau \sim 5400$ sec, but increasing $\tau$ increases the precision, thus $\tau \sim 1$ sec can be taken as a lower limit.} $\tau \sim 1$ sec, we find $S_y (f) = h_0 \sim 2.42 \times 10^{-26}$. The square root of the spectral density gives a measure of the accuracy to which the frequency can be measured. For the numbers above, we have $\sqrt{S_y (f)/\tau} \sim 1.56 \times 10^{-13}$, and this should be compared to $\frac{\Delta f}{f} = \frac{\Delta E}{{\tilde E}_i}$, where $\Delta f$ and $\Delta E$ are the frequency and energy shift of the side bands, respectively, and $f$ and ${\tilde E}_i$ are the central frequency and energy, respectively. Using the central frequency reference \cite{laurent} gives $f = \frac{{\tilde E}_i}{h} = 9.19 \times 10^9$ Hz, and from \eqref{energy-3} we have $\Delta f = \frac{\Delta E}{h} = \frac{2.4 \times 10^{-7} {\rm eV}}{4.14 \times 10^{-15} {\rm eV Hz ^{-1}}} \sim 58 {\rm MHz}$. Thus $\frac{\Delta f}{f} \sim 6.31 \times 10^{-3}$. Comparing this with the square root of the spectral density, we have $\frac{\Delta f}{f} \sim 6.31 \times 10^{-3} \gg  1.56 \times 10^{-13} \sim \sqrt{S_y (f)/\tau}$. These side bands should be easily observable via the ACES.

One final point is that we need to take into account that the weighting of the maximum side band, $J_\alpha (\alpha)$, decreases as $\alpha$ increases, as shown in Fig. 1. For the atomic case with $\alpha \sim 10^{11}$ we find that $J_\alpha (\alpha) \sim 10^{-4}$. This decrease in the weighting of the maximum side band will decrease the expected signal, but as long as the signal to noise ratio (SNR) is greater than 1, the side band should be observable. The measure of SNR is $\frac{J_\alpha (\alpha)}{\sqrt{S_y (f)/\tau}}$. For the parameters above, $\sqrt{S_y (f)/\tau} \sim 1.56 \times 10^{-13}$ and $J_\alpha (\alpha) \sim 10^{-4}$, we have $\frac{J_\alpha (\alpha)}{\sqrt{S_y (f)/\tau}} \approx 6.4 \times 10^8 \gg 1$. Thus, even given the reduction due to the weighting, $J_\alpha (\alpha)$, the side band should be observable. 

\subsection{Nuclear System: M{\"o}ssbauer Effect} -- Another way to test for the gravitationally induced side bands from  \eqref{energy-4} is the M{\"o}ssbauer effect \cite{moss}. The M{\"o}ssbauer effect involves the emission and absorption of gamma rays. The recoil due to the emission/absorption of the gamma ray is taken up by the entire lattice of the material, so that there is effectively no recoil. This makes the emissions/absorption lines very narrow, approaching the limit set by the uncertainty principle. For $^{57}$Fe with a lifetime of $\tau \sim 10^{-8}$ sec, and the emission of a 14.4 keV photon, one has a width of $\Delta E = \frac{\hbar}{\tau} \approx 10^{-8}$ eV. This is much smaller than the shift of the side bands of $\sim \pm 10^{-4}$ eV, thus allowing one to observe the side bands.

The experimental set-up would be to have a M{\"o}ssbauer $^{57}$Fe spectrometer inside a satellite that is in low-Earth, almost circular orbit, and look for the predicted side bands. The M{\"o}ssbauer effect is ideally suited to observe fine spectral details of this type. For example, in \cite{kistner} the M{\"o}ssbauer effect was used to determine the hyperfine structure of $^{57}$Fe which had splitting smaller than the predicted $\pm 10^{-4}$ eV for the gravitational AB effect. 

This proposed use of the M{\"o}ssbauer effect can be compared to the Pound-Rebka experiment \cite{pound}. In the Pound-Rebka experiment, the very small red-shift/blue-shift of photons rising/falling vertically in the Earth's gravitational field was measured using the M{\"o}ssbauer effect. The gamma-ray emitter/absorber used was $^{57}$Fe, and the emitter/absorber were spatially separated by about 20 meters vertically, so that as the photons rose/fell in going from emitter to absorber, they would be gravitationally red/blue shifted by a very small amount, which was detectable due to the high precision of M{\"o}ssbauer spectroscopy.

In the current paper, the proposed use of the M{\"o}ssbauer effect is a temporal version of the Pound-Rebka experiment. Instead of having the emitter and absorber separated spatially at different gravitational potentials, the emitter and absorber will be placed very near each other, preferably perpendicular to the gravitational field, since we want the emission and absorption to occur at the same gravitational potential spatially. However now the emitter and absorber will be temporally separated, so as the satellite orbits, the energy spectrum of the $^{57}$Fe nucleus should develop the side bands predicted in  \eqref{energy-4} which should be detectable given the precision of the M{\"o}ssbauer effect. 

Using atomic clocks like those of the ACES system has the advantages that this system should be operational in the short term -- one or two years \cite{aces} -- and $\alpha _{atomic } > \alpha _{nuclear}$ by about three orders of magnitude which means that the weighting for the maximum side band will be larger in the atomic clock system as compared to the M{\"o}ssbauer set up. The M{\"o}ssbauer set up has the advantage that the splitting of the side bands is larger by about three orders of magnitude compared to the atomic clock set up. Both set ups appear viable ways to test for this version of the gravitational AB effect.    

\section{Conclusions} -- In this work we propose a novel approach to testing the gravitational AB effect, which is distinct from the detection of the gravitational AB effect in \cite{overstreet}. The work by Overstreet {\it et al.} used the standard experimental signature for the gravitational AB effect: finding a relative phase shift between two beams of particles ($^{87}$Rb atoms) which were split along different paths and passed through different gravitational potentials. This observation of a phase shift is also the usual way in which the electromagnetic AB effect is observed. Here we propose placing a quantum system in a time varying gravitational potential and looking for the appearance of energy side bands. This is the gravitational version of the proposal in \cite{chiao-2023} to probe the scalar electric AB effect. Due to the weakness of the gravitational interaction, in order to get energy side bands large enough to observe, we need the time variation of an astrophysical large mass. This can be achieved by placing the quantum system in a satellite in a low Earth, almost circular orbit. The slight change in the gravitational potential between apogee and perigee provides the change in gravitational potential. Also, since the satellite is in free fall, this effectively eliminates the gravitational field, locally, via the equivalence principle. 

Starting with \eqref{tdse}, we carried out an analysis parallel to the one used for the scalar-electric AB effect \cite{chiao-2023} but applied it to the gravitational case. As in the scalar-electric case, we found that the energy levels of the quantum system developed energy side bands as given in \eqref{energy-3} and \eqref{energy-4} for atomic and nuclear systems, respectively. In the scalar-electric case it was relatively easy to change the size of the electric potential, $V_0$, and the frequency of the changing electric potential, $\Omega$, over a wide range. In contrast, for the gravitational case this was not possible since the variation of the gravitational potential and the frequency are controlled by the parameters of the satellite orbit, which has a much more narrow range as compared to the scalar-electric case. The one parameter which we could change in the gravitational AB case is whether our quantum system is atomic or nuclear leading to $m$ being the electron mass or nucleon mass, respectively. 

The gravitational side bands generated are much smaller that in the scalar-electric case, as expected. For the atomic case, the sides bands are $\sim \pm 10^{-7}$ eV, and  for the nuclear case, the sides bands are $\sim \pm 10^{-4}$ eV. Nevertheless, these small shifts can be observed using precision spectroscopy of atomic clocks in the optical/microwave frequency range ({\it e.g.} like those of ACES) or the  M{\"o}ssbauer effect. In fact, the ACES program \cite{aces} might be able to test this in the near future. \\~\\

{\bf Acknowledgements}: MET is funded by the ARC Centre of Excellence for 474 Engineered Quantum Systems, Grant No. CE170100009 475 and the ARC Centre of Excellence for Dark Matter Particle 476 Physics, Grant No. CE200100008.

\end{document}